\documentstyle[preprint,aps]{revtex}                                            
\begin{document}                                                                
\preprint{DOE/ER/41014-1-N97}                                                   
\draft                                                                          
\title{NEW APPROACH TO $^4$He CHARGE DISTRIBUTION}                              
\author{L. Wilets, M. A. Alberg\footnote{permanent address: Department of       
Physics, Seattle University, Seattle WA 98122}}                                 
\address{Department of Physics, Box 351560, University of Washington,           
 Seattle, WA 98195-1560, USA}                                                   
\author{S. Pepin and Fl. Stancu}                                                
\address{Universit\'{e} de Li\`ege, Institut de Physique B.5, Sart-Tilman,      
 B-4000 Li\`ege 1, Belgium}                                                     
\author{J. Carlson}                                                             
\address{Theoretical Division, Los Alamos National Laboratory, Los Alamos,      
 NM 87545, USA}                                                                 
\author{W. Koepf}                                                               
\address{Department of Physics, The Ohio State University, Columbus OH 43210,   
 USA}                                                                           
\maketitle                                                                      
\begin{abstract}                                                                
We present a study of the $^4$He charge distribution based on
realistic          nucleonic wave functions and incorporation of the
nucleons'                      quark substructure. The central depression of
the proton point density seen in modern four-body calculations is too small by
itself to lead to a correct description of the charge distribution.  We
utilize six-quark structures calculated in the Chromodielectric Model for N-N
interactions, and we find a swelling of the proton charge distribution as the
internucleon                                                             
distance decreases. These charge distributions are combined with
the           
$^4$He wave function using the Independent Pair Approximation and two-body      
distributions generated from Green's Function Monte Carlo calculations.  We
obtain a reasonably good fit to the experimental charge distribution without
including meson exchange
currents.                                                        
\end{abstract}                                                                  
\section{INTRODUCTION}                                                          
                                                                                
The charge distribution of nuclei has been the subject of experimental          
studies for more than forty years.  Electron scattering and muonic atoms        
now provide detailed descriptions of the full range of stable, and many         
unstable nuclides. Unique among the nuclides are the isotopes $^3$He and        
$^4$He because they exhibit a central density about twice that of any           
other nuclei. There is a long-standing apparent discrepancy between the         
experimentally extracted charge distributions and detailed theoretical          
structure calculations which include only nucleon degrees of freedom.           
                                                                                
McCarthy, Sick and Whitney\cite{mc,sick} performed electron scattering          
experiments on these isotopes up to momentum transfers of 4.5 fm$^{-1}$         
yielding a spatial resolution of $~0.3$ fm.  They extracted a ``model           
independent" charge distribution, which means that their analysis of the data   
is not based upon any assumed functional form for the charge distributions.     
Their results are shown in Fig. 1.  Taken alone, they  do not                   
appear                                                                          
to be extraordinary.  However, using the experimentally measured proton         
form factor, which has an rms radius of about 0.83 fm, they unfolded the        
proton structure from the charge distributions to obtain proton point           
distributions.  For both isotopes there is a significant                        
central depression of about 30\% extending to about 0.8 fm. Sick\cite{sick}     
also obtained results where relativistic and meson effects                      
are included.  These are shown for $^4$He in Fig. 2. One note of caution        
here is that it is not possible to subtract these effects from the              
experimental data in a completely model-independent way.                        
                                                                                
One might assume that such a central depression is to be expected because       
of the short-range repulsion of the nucleon-nucleon interaction.  This is       
not borne out by numerous detailed theoretical calculations (see, for           
example, Ref.\cite{friar})  none of which                                       
finds a {\it significant} central depression, certainly not of the above        
magnitude. Relatively smaller central depressions are found in                  
Green's Function Monte Carlo
(GFMC) calculations of $^4$He
for realistic models of the two- and three-nucleon
interaction \cite{wiringa,carlson1}.                                       
                                                                                
The status of theoretical structure calculations through mass number 4 is       
very satisfactory at present.  Given any assumed interaction, the few body      
problem can be solved to within tenths of an MeV in energy and the wave         
function can be calculated to a precision better than that required for the     
present discussion.                                                             
                                                                                
In using a nucleonic wave function to construct a charge distribution, one      
must use an assumed nucleon charge density and the possibility of meson         
exchange contributions.                                                         
While the meson exchange contributions in the transverse channel are            
well-constrained (at least at moderate momentum transfer) by current            
conservation, no such constraint is available in the longitudinal               
channel. Indeed, meson exchange contributions are of relativistic order         
and hence one must be careful when interpreting them with non-relativistic      
wave functions.                                                                 
                                                                                
Given these caveats, it is possible to reproduce reasonably well the            
longitudinal form factors of three- and four-body nuclei within a               
nucleon-plus-meson-exchange model \cite{wiringa,carlson1,schiavilla}. The
current and charge operators are constructed from the N-N interaction
and required to satisfy current conservation at non-relativistic order.
The resulting meson-nucleon form factors are quite hard, essentially
point-like               
\cite{wiringa,carlson1}. This raises the possibility of explaining the
form factors in quark or soliton based models, which would
describe the short-range two-body structure of the nucleons in a
more direct way than is available through meson exchange current
models. See, for example, the model by Kisslinger et al.
\cite{kis}.                                      
                                                                                
We present here a possible explanation of the electric form factors which       
is consistent                                                        
with theoretical few body calculations.  It involves the variation of the       
proton charge form factor (size) as a function of                               
proton-nucleon separation. This is not depicted as an average `swelling'        
of the nucleon, but as a result of short-range dynamics in the                  
proton-nucleon system, as discussed in the next section.                        
We relate the variation of the proton size                                      
to the quarks dynamics, here described by the Chromodielectric Soliton Model
(CDM).   
                                                                                
\section{QUARK SUBSTRUCTURE OF NUCLEI AND NUCLEONS}                             
                                                                                
Within the context of soliton models, there have been numerous                  
calculations of nucleon size in nuclear media.  Most of these                   
involve immersion of solitons in a uniform (mean) field generated               
by other nucleons \cite{birse}.  Another approach has been the                  
3-quark/6-quark/9-quark bag models, which has been applied to various           
nuclear properties, including the EMC effect \cite{pirner}.                     
A hybrid quark-hadron model has been applied by Kisslinger  et                  
al.\cite{kis} to the He electric form factors with some                         
success.                                                                        
                                                                                
In a series of papers, Koepf, Pepin, Stancu and                                 
Wilets\cite{statics,emc,dynamics} have studied the                              
6-quark substructure of the two-nucleon problem in the framework of the         
chromodielectric soliton model \cite{book}. In particular, they calculated      
the variation of the quark wave functions with inter-nucleon separation.        
Contrary to previous expectations, the united 6-quark cluster does not          
exhibit a significant decrease in the quark momentum in spite of an             
increase in the volume available to the individual quarks \cite{emc}.           
This is due to                                                                  
configuration mixing of higher quark states.   Such a momentum                  
decrease had been proffered as an explanation of the EMC effect \cite{blank}.   
However, the united cluster does have approximately twice the                   
confinement volume of each 3-quark                                              
cluster, and the quarks extend to a volume nearly three times that of the       
3-quark clusters, again enhanced by configuration mixing of excited states.     
                                                                                
In Fig. 3 we exhibit the proton form factor based on the calculations of        
Pepin {\it et al.}\cite{dynamics} extracted as follows: the soliton-quark       
structure is a 6-quark deformed                                                 
composite; the Generator Coordinate Method was used                             
to build a dynamical nucleon-nucleon potential; it involves a Fujiwara          
transformation, which relates the deformation of the six-quark bag to           
the effective nucleon-nucleon separation $r_{NN}$.                              
 The proton rms radius is then defined to be                                    
\begin{equation}                                                                
r_p=\sqrt{<r^2>-r_{NN}^2/4}                                                     
\end{equation}                                                                  
where the quark density $\rho_q$ used in calculating                            
$<r^2>=\int \rho_q r^2 d^3r$ has been obtained from 6-quark CDM calculations    
[13] and normalized to unity.                                                   
For separated solitons, the $r_{NN}$ is just the separation of the soliton      
centers and $r_p=0.83$ fm as indicated by the horizontal line labeled ``Free    
Proton".  Large deformations (near separation) are difficult to                 
calculate. Shown also in the figure is a Gaussian approximation                 
(dashed line) fitted to the CDM result at $r_{NN}=0$, $r_{NN}=1$ fm, and        
in the asymptotic region.                                                       
                                                                                
To obtain the latter approximation,                                            
we assume a Gaussian form for the variable ``proton" charge density.   Then     
the charge distribution due to two nucleons is                                  
\begin{eqnarray}                                                                
\rho_{pair}(\bbox{r_i,r_j;r}) =  \left\{ \delta_{ip}                            
\exp\big[-|\bbox{r-r_i}|^2/b^2(r_{ij})\big] \right. \nonumber \\                
 \left. +\delta_{jp}                                                            
\exp\big[-|\bbox{r-r_j}|^2/b^2(r_{ij})\big] \right\} /                          
\pi^{3/2}b^3(r_{ij})                                                            
\end{eqnarray}                                                                  
where we indicate explicitly that $b$ is a function of the                      
distance $r_{ij}$ between the nucleons $i$ and $j$.  Here ``$p$" stands
for                                                                         
``proton" and the Kronecker symbols pick out protons among $i$ and
$j$.         
                                                                                
We allow the Gaussian size $b$ to depend on the internucleon distance           
according to                                                                    
\begin{equation}                                                                
b(r')=b_0 [1+Ae^{-r'^2/s^2}]\,,                                                 
\end{equation}                                                                  
with the free proton value, $b_0=\sqrt{2/3}$ 0.83 fm.                           
The Gaussian fit shown in Fig. 3 corresponds to $A=0.45$ and $s=1.92$ fm.       
                                                                                
 Using the Independent Pair Approximation (IPA) and                             
Eq. (2), we can calculate the charge distribution from the above result by      
employing a two-body correlation function,                                      
$\rho_2(\bbox{r_i,r_j})$,                                                       
\begin{equation}                                                                
\rho_{ch}(r)=\sum_{i<j}\int d^3r_i\int                                          
d^3r_j\,\rho_2(\bbox{r_i,r_j})                                                  
\rho_{pair}(\bbox{r_i,r_j;r)}/3 .                                               
\end{equation}                                                                  
There are six pairs (i,j) and each ``proton" appears three times, hence         
the factor 1/3.                                                                 
The two-body correlation function is generated from the nuclear wave function
(described in the next section) by taking the
average                                                              
\begin{equation}                                                                
\rho_2(\bbox{r_1,r_2})=\int|\psi(\bbox{r_1,r_2,r_3,r_4})|^2d^3r_3\,d^3r_4\,,    
\end{equation}                                                                  
with the constraint
$\bbox{r_1+r_2+r_3+r_4=0}$.                                                     
                                                                                
\section{$^4$He CHARGE DISTRIBUTION : RESULTS}
                                                                     
The $^4$He wave function $\psi(\bbox{r_1,r_2,r_3,r_4})$ has been obtained
by solving the non-relativistic Schr\"odinger equation :
\begin{equation}
\left[ \sum_{i} - \frac{\hbar^2}{2 m} \nabla_{i}^{2} + \sum_{i<j} V_{ij} +
\sum_{i<j<k} V_{ijk} \right] \psi = E \psi .
\label{sch}
\end{equation}
where $V_{ij}$ and $V_{ijk}$ are respectively the $v_{18}$ Argonne
nucleon-nucleon interaction and the Urbana 9 three-nucleon interaction. 
The parameters entering $V_{ij}$ have been fitted to the deuteron and
two-nucleon scattering data. The parameters of $V_{ijk}$ were determined by
fitting the binding energy of $A=3$ nuclei. The Schr\"odinger equation
(\ref{sch}) has been solved by using the Green's Function Monte Carlo
(GFMC) method, which has proven to be very valuable in studying light nuclei,
and produced more accurate results than the so-called Variational Monte Carlo
(VMC) method. A typical difference, important for the present study, is that
within 0.5 fm of the centre-of-mass the GFMC point density has a slight hole
which does not appear in VMC results. Details about the GFMC method and the
form of the interaction potentials used are reviewed in Ref.\cite{carlson1}. 
The combination of the potentials $V_{ij}$ and $V_{ijk}$ introduced above
gives the correct binding energy and approximately the correct rms radius for
$^4$He.  Previous calculations of other properties of $^4$He
have been done with an older nucleon-nucleon
interaction                                                     
$v_{14}$ \cite{carlson88}.
  
We have calculated the charge distribution for $^4$He in the spirit of
the Independent Pair Approximation using Eq. (4).
The proton size parameter        
$b(\bbox{r_1-r_2})$ given by Eq. (3) was taken from the Gaussian fit to the     
calculations of Pepin {\it et al.}\cite{dynamics} presented in Fig. (3).        
The resulting  distribution is shown in Fig. (4) (solid line).                  
The improvement over the free proton case, i.e., a Gaussian with a              
constant $b=b_0$ (dotted line) is impressive and leads to                       
a fairly good agreement with the data.  In the same figure we also indicate     
the point density (dashed line) derived straightforwardly from the $^4$He wave  
function described above. Even if this
density shows a central depression, it is not sufficient by
itself to reproduce the data, when combined with the free
proton form factor.

\section{CONCLUSIONS}                                                           
                                                                                
We succeeded in reproducing fairly well the $^4$He charge distribution          
by assuming a proton size which increases with increasing density.              
                                                                                
We have identified the origin of the variable proton size through the           
structure function obtained from dynamical 6-quark N-N studies in the           
spirit of the Independent Pair Approximation.                                   
                                                                                
We could probably improve our results if we recalculate meson effects using     
the quark structure functions given (say) by the six-quark IPA model.  This     
item is a topic for further investigation.                                      
                                                                                
In addition, one                                                                
must study the predictions of such models for quasi-elastic scattering. In      
the quasi-free regime, nucleons models produce a good description of the        
data as long as realistic nucleon interactions, including charge exchange,      
are incorporated in the final-state interaction \cite{carlson}. Unlike the      
charge form factor, two-body charge operators are expected to play a much       
smaller role here, principally because this is the dominant channel and         
there is little interference. The combination of the two regimes provides a     
critical test for models of structure and dynamics in light nuclei.             
                                                                                
\acknowledgments                                                                
                                                                                
We wish to thank C. Horowitz and others for valuable discussions.               
This work is supported in part by the U. S. Department of                       
Energy and by the National Science Foundation.

\begin{figure}                                                                  
\noindent\caption{Model-independent charge distributions for $^3$He and         
$^4$He extracted                                                                
from experiment.  Reproduced from McCarthy {\it et al.}[1].                    
, who state that ``the extreme limits of $\rho(r)$ cover the
statistical, systematical as well as the completeness error of the
data."                               
\label{fig1}}                                                                   
                                                                                
\bigskip                                                                        
\bigskip                                                                        
                                                                                
\noindent \caption{Point-proton density distribution for $^4$He  obtained by    
unfolding the free proton form factor, allowing for meson exchange              
corrections and relativistic effects.  Reproduced from Sick[2].                 
\label{fig2}}                                                                   
                                                                                
\bigskip                                                                        
\bigskip                                                                        
                                                                                
        \caption{Proton rms charge radius $r_p$ of Eq. (1) as a function        
of inter-nucleon separation.  The line labeled CDM is the calculated            
chromodielectric model result.                                                  
The dashed line is a Gaussian approximation, normalized to the                  
free value, with a size parameter given by Eq. (3)}                             
        \label{Figure_3}                                                        
                                                                                
\bigskip                                                                        
\bigskip                                                                        
                                                                                
        \caption{$^4$He density distributions: The dashed line is the point     
density from a parameterized Green's
Function Monte Carlo calculation. The
curve labeled ``free proton" is the charge
distribution obtained from a Gaussian proton charge                  
distribution with a fixed size parameter (as is usually done). 
The curve labeled ``variable proton size" uses the
Gaussian fit of Fig. 3. We also indicate half the
normal nuclear density 0.17/2          
fm$^{-3}$.}                                                                     
\label{Figure_4}                                                                
                                                                                
\end{figure}                                                                    
                                                                                
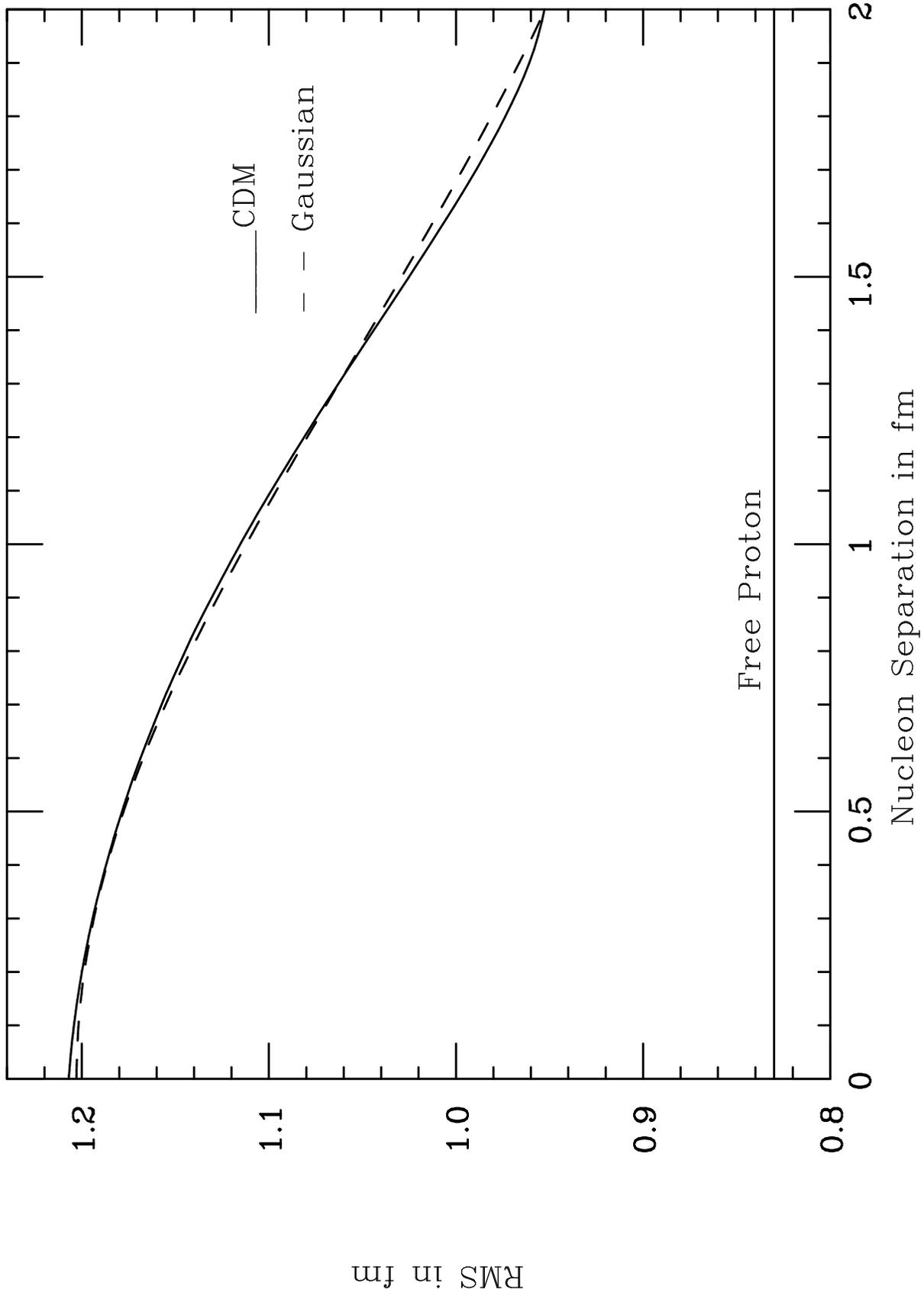
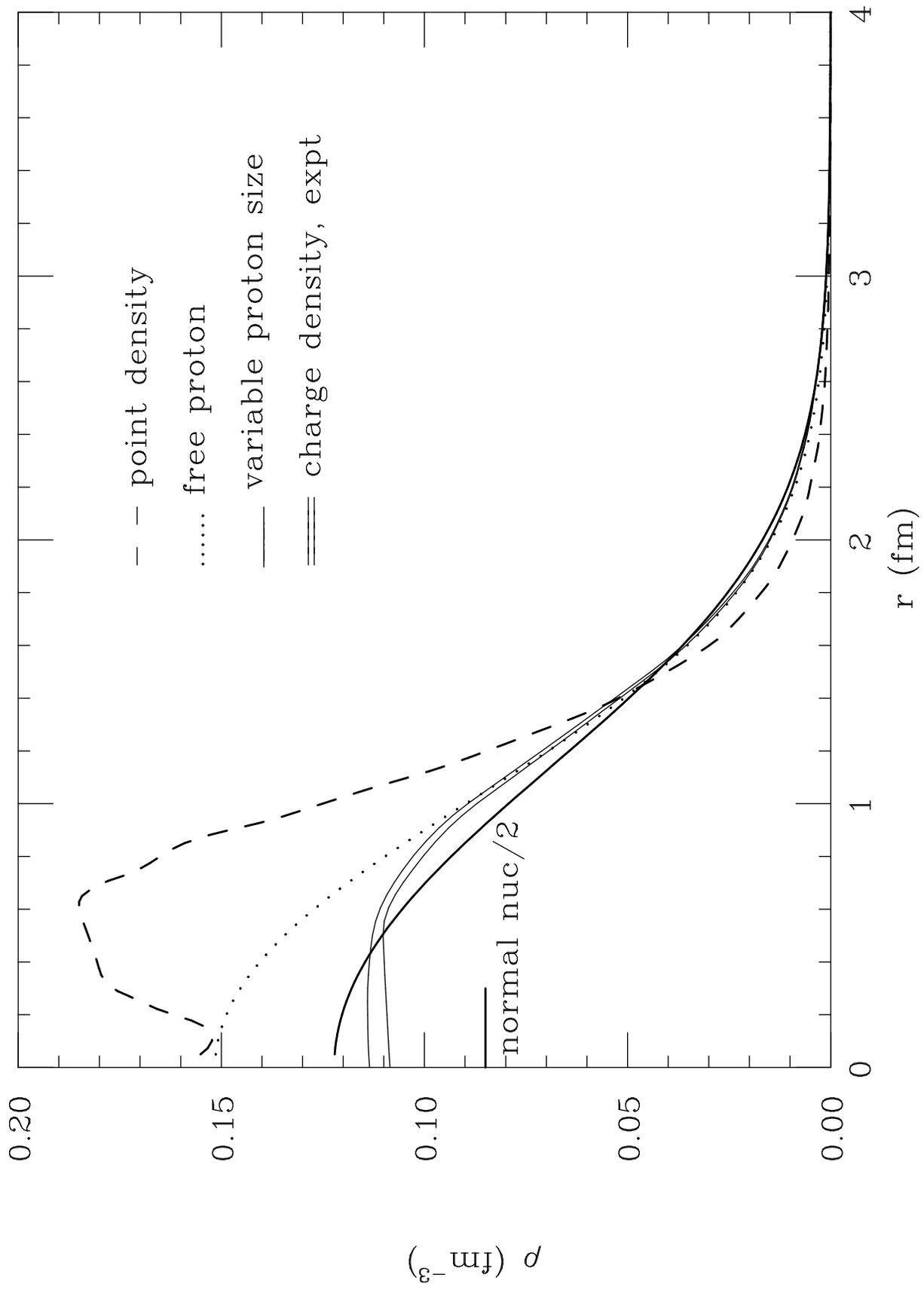
\end{document}